# Stability and low-energy orientations of interphase boundaries in multiaxial ferroelectrics: phase-field simulations

Yang Zhang,[1,2,*] Fei Xue,[3,†] Bo Wang,[3] Jia-Mian Hu,[4] Shuai Dong,[1] Jun-Ming Liu,[2] Long-Qing Chen[3]

[1]*School of Physics, Southeast University, Nanjing 211189, China*

[2]*Laboratory of Solid State Microstructures and Innovation Center of Advanced Microstructures, Nanjing University, Nanjing 210093, China*

[3]*Department of Materials Science and Engineering, The Pennsylvania State University, University Park, PA 16802, USA*

[4]*Department of Materials Science and Engineering, University of Wisconsin-Madison, Madison, WI 53706, USA*

---

[*] E-mail address: zhangyang919@gmail.com.

[†] E-mail address: xuefei5376@gmail.com.

**Abstract**

The coexistence of different ferroelectric phases enables the tunability of the macroscopic properties and extensive applications from piezoelectric transducers to non-volatile memories. Here we develop a thermodynamic model to predict the stability and low-energy orientations of boundaries between different phases in ferroelectrics. Taking lead zirconate titanate and bismuth ferrite as two examples, we demonstrate that the low-energy orientations of interphase boundaries are largely determined by minimizing the electrostatic and elastic energies. Phase-field simulations are employed to analyze the competition between the interfacial energy and the electrostatic and elastic energies. Our simulation results demonstrate that the lowering of crystal symmetry could occur due to the electrical and mechanical incompatibilities between the two phases, which can be used to explain the experimentally observed low-symmetry phases near morphotropic phase boundaries. Our work provides theoretical foundations for understanding and controlling the interphase boundaries in ferroelectric materials for multifunctional applications.

# I. Introduction

Ferroelectric (FE) materials have been widely applied in memories, piezoelectric micro-components, high-frequency electronics, etc. [1,2], while the rapid development of nanotechnology provides new challenges and opportunities. Driven by the trend of device miniaturization and multifunctionalization, extensive efforts have been made in the design of micro/composite structures, the modulation of couplings among multiple degrees of freedom, the control of nanodomains and their boundaries, etc. [3–7]. Among them, the manipulation of FE domain walls (DWs, the interfaces between two domain variants of the same phase) and interphase boundaries (IBs, the interfaces between two domains of different phases) has been proven particularly useful for obtaining unprecedented properties and achieving enhanced performances, offering complicated underlying physics as well as plentiful applications [8,9]. For instance, the profiles of DWs may influence the conduction of FE materials [10,11], and different resistance states could be manipulated via electric fields [9,12]. The engineered IBs in FE thin films can exhibit excellent energy storage efficiency and thus extend the applications of dielectric capacitors [13]. Furthermore, the crystal structure near the morphotropic phase boundaries in FE solid solutions such as lead zirconate titanate (PZT), lead magnesium niobate-lead titanate (PMN-PT), and potassium sodium niobate alters sharply, and the phases will transform to each other along with the small alteration of chemical compositions or mechanical stresses [14]. In this case, the easy movement of the IBs under external stimuli is often associated with giant piezoelectricity [15–17], and thus a deeper understanding of these boundaries is necessary.

From an energy perspective, the stability of DWs and IBs are determined by the interplay of various energy contributions including the short-range interfacial energy and the long-range elastic and electrostatic energies [18,19]. Ideally, the single domain should be the ground state for an infinite FE bulk crystal under zero external stress and electric field, since extra local energy would arise in the presence of a domain interface. Nevertheless, the formation of polydomain structures is usually a spontaneous process to lower the overall elastic and/or electrostatic energy arising from the mechanical and/or electrical boundary conditions for finite-size ferroelectrics. For example, a mono-domain FE thin film with uniform out-of-plane polarization can accumulate bound charges on the surfaces if not properly screened, which can induce large depolarization field across the film. To stabilize such out-of-plane ferroelectricity, 180-degree DWs tend to form spontaneously to lower the electrostatic energy [20–23]. Similarly, in FE thin films subject to the clamping effect from substrates, ferroelastic domains with non-180-degree DWs or IBs emerge to release the otherwise large elastic energy associated with a coherent mono-domain state [24].

In general, the orientations of low-energy domain interfaces are not arbitrary. It is known that a stable DW should satisfy the electrical compatibility condition (i.e., polarization continuity) and the mechanical compatibility condition at the same time [25], which guarantee that both the polarization vectors and the spontaneous strain tensors are continuous across the corresponding wall (see Eqs. (S9) and (S10) in Section SI of the Supplemental Material [26]). In fact, systematic analyses of these two compatibility conditions have been well established in literature [25,27–30]. For an IB, however, the mechanical compatibility condition generally cannot be satisfied, and it remains elusive how to theoretically analyze its stability and low-energy orientations.

In this work, we quantify the energy changes due to the presence of DWs or IBs in ferroelectrics based on an analytical model, incorporating the long-range electrostatic and elastic energies. In Section II, we establish the model by introducing two incompatibility factors to evaluate the degrees of electrical and mechanical incompatibilities and considering the contributions from the phase separation process. In Section III, taking lead zirconate titanate and bismuth ferrite as two examples, we calculate the low-energy orientations of DWs and IBs based on the minimization of the two incompatibility factors. The competition between the short-range and long-range interactions is demonstrated by using phase-field simulations. The total excess energy as a function of the domain width is discussed, which predicts a critical domain width for the stability of domain interfaces. We also demonstrate that the crystal symmetry of highly strained $BiFeO_3$ thin films can be lowered due to the electrical and mechanical incompatibilities between the two phases. Section IV concludes our findings.

## II. Thermodynamic analysis of compatible and incompatible domain walls and interphase boundaries

The formation of DWs or IBs in FE materials is accompanied by a local excess energy called the interfacial energy, which is commonly characterized by its surface density denoted as $\gamma_{DI}$. The interfacial energy origins from the discontinuity of order parameters across the DWs or IBs which causes extra Landau free energy and gradient energy. Apart from this short-range interaction, the formation of DWs or IBs may also affect the long-range electrostatic and elastic energies.

The electrostatic energy density can be written as

$$f_{\text{elec}} = -\frac{1}{2}\varepsilon_0 \kappa_{ij}^{\text{b}} E_i E_j - P_i E_i, \tag{1}$$

where $\varepsilon_0$ is the vacuum permittivity, $\kappa_{ij}^{\text{b}}$ is the background permittivity tensor [31,32], $E_i$ is the electric field component, and $P_i$ is the spontaneous polarization component (see Section SI of the Supplemental Material [26]). The Einstein convention is used throughout this work. The elastic energy under stress-controlled boundary conditions is written as

$$f_{\text{elas}} = -\frac{1}{2} s_{ijkl} \sigma_{ij} \sigma_{kl} - \sigma_{ij} \varepsilon_{ij}^0, \tag{2}$$

where $s_{ijkl}$ is the fourth-rank compliance tensor, $\sigma_{ij}$ and $\varepsilon_{ij}^0$ are the components of stress and eigenstrain, respectively.

For simplicity, we consider a one-dimensional periodic system consisting of two alternating domains which can either represent two domain variants of the same phase or two phases. The interface plane is denoted by its unit normal vector **n**. As shown in Fig. 1(a), the system with the period $L$ can be divided into Domain I and Domain II with fraction $w^{\text{I}}$ and $w^{\text{II}}$, respectively. Apparently, $w^{\text{I}} + w^{\text{II}} = 1$. For Domain $p$ ($p$ = I, II), the polarization vector is $\mathbf{P}^p$, and the eigenstrain tensor is $\boldsymbol{\varepsilon}^{0(p)}$. In addition, the sharp interface assumption is employed in the analytical derivation.

In most cases of DWs, the mechanical compatibility condition [Eq. (S10)] has two roots which represent two perpendicular DWs, and one of them satisfies the electrical compatibility condition [Eq. (S9)] at the same time. For other cases, there exist either infinite permissible wall orientations or zero mechanically compatible DW [8].

As mentioned in Section I, the mechanical compatibility condition can be satisfied only for homophase DWs but has no solutions for heterophase IBs in general, and extra electrostatic and/or elastic energies would be gained for incompatible interfaces. From Eqs. (1) and (2), we can obtain the average electrostatic and elastic energy densities of the polydomain system as well as that of the mono-domain states. The difference of electrostatic or elastic energy between the polydomain structure and two corresponding mono-domain ones measures the long-range influence of DWs or IBs. After some derivations (see Section SII of the Supplemental Material [26]), the electrostatic energy density change due to the introduction of DWs or IBs is

$$\Delta f_{\text{elec}} = \frac{1}{2} w^{\text{I}} w^{\text{II}} \frac{\left(\mathbf{P}^{\Delta} \cdot \mathbf{n}\right)^2}{\varepsilon_0 \kappa_{\mathbf{n}}^{b}} := \frac{1}{2} w^{\text{I}} w^{\text{II}} Q(\mathbf{n}), \tag{3}$$

where $\mathbf{P}^{\Delta} = \mathbf{P}^{\text{I}} - \mathbf{P}^{\text{II}}$ and $\kappa_{\mathbf{n}}^{b} = \kappa_{ij}^{b} n_i n_j$. Here, we define the so-call *electrical incompatibility factor* as

$$Q(\mathbf{n}) = \frac{\left(\mathbf{P}^{\Delta} \cdot \mathbf{n}\right)^2}{\varepsilon_0 \kappa_{\mathbf{n}}^{b}}, \tag{4}$$

which depends on the interface normal $\mathbf{n}$ for the two given FE domains. Clearly, $Q(\mathbf{n}) \geq 0$. If $Q(\mathbf{n})$ is zero, i.e., $\mathbf{P}^{\Delta} \cdot \mathbf{n} = 0$ [Eq. (S3)], the interface is electrically compatible.

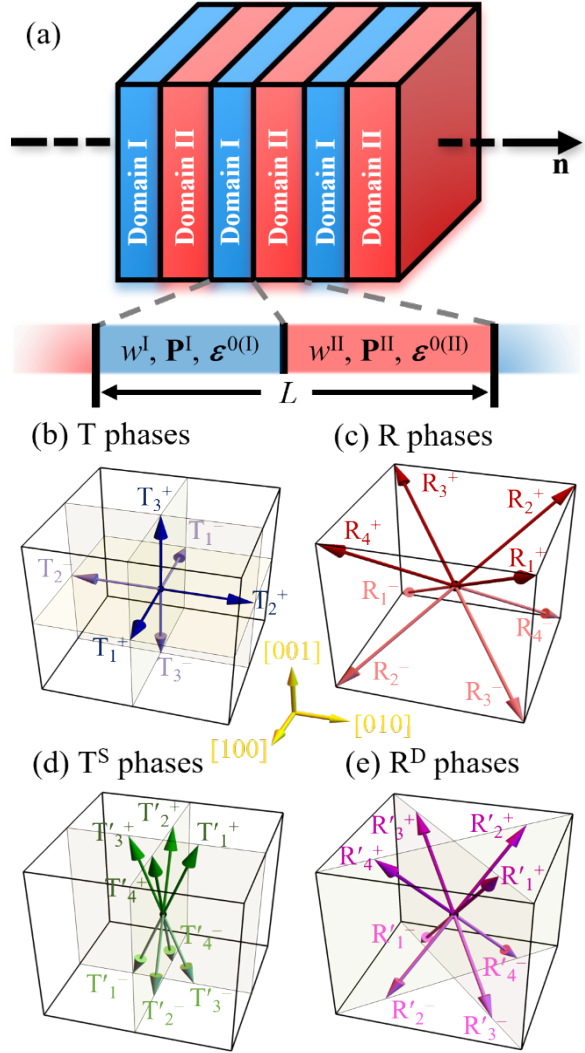

FIG. 1. (a) The sketch of the studied system with alternating two domains. (b-c) Polarization directions of the domain variants in lead zirconate titanate (PZT) and bismuth ferrite (BFO) single crystals, including the tetragonal (T) phase and the rhombohedral (R) phase. (d-e) Polarization directions of the domain variants in BFO thin films, including the super-tetragonal ($T^S$) phase and the distorted rhombohedral ($R^D$) phase.

Similarly, the excess elastic energy density due to the formation of DWs or IBs can be

written as

$$\Delta f_{elas} = \frac{1}{2} w^{I} w^{II} B(\mathbf{n}). \tag{5}$$

Here the *mechanical incompatibility factor* is defined as

$$B(\mathbf{n}) = c_{ijkl}\varepsilon_{ij}^{\Delta}\varepsilon_{kl}^{\Delta} - n_i\sigma_{ij}^{\Delta}\Omega_{jk}\sigma_{kl}^{\Delta}n_l, \tag{6}$$

where $c_{ijkl}$ is the elastic stiffness tensor, $\sigma_{ij}^{\Delta} = c_{ijkl}\varepsilon_{kl}^{\Delta}$ with $\varepsilon_{ij}^{\Delta} = \varepsilon_{ij}^{0(I)} - \varepsilon_{ij}^{0(II)}$, and $\Omega$ is a second-rank tensor calculated by $\Omega_{ik}^{-1} = c_{ijkl}n_j n_l$ (see Section SIII of the Supplemental Material [26] for detailed derivations). For two given ferroelastic domains, or two given eigenstrains, $B(\mathbf{n})$ only depends on the interface normal $\mathbf{n}$. It is noteworthy that the form of $B(\mathbf{n})$ can also be derived from Khachaturyan's microelasticity theory [33]. Generally, we have $B(\mathbf{n}) \geq 0$, and the mechanical compatibility condition [Eq. (2)] is the special case as $B(\mathbf{n}) = 0$.

In addition, the phase separation process of a homogeneous phase into a mixture of two or more phases is one of the origins for the formation of polydomain structures. The most known example is the spinodal decomposition [34,35]. Also, in FE thin films, the strain phase separation [36,37] from one strain state to different strain states of structural domain/phase variants has been commonly observed in many systems. The energy reduction owing to the phase separation $\Delta f^{de}$ can be obtained by polynomial fittings to the curves of free energy versus corresponding separated variables and then calculating their common tangents [36] (see Section SIV of the Supplemental Material [26] for a simple example). It should be noted that $\Delta f^{de}$ is independent of the interface normal $\mathbf{n}$.

We can see that several contributions are associated with the introduction of DWs or IBs

$\frac{1}{2}w^{\mathrm{I}}w^{\mathrm{II}}[Q(\mathbf{n})+B(\mathbf{n})]$ from electrical and mechanical incompatibilities depending on the fraction of each domain and the direction of the domain interfaces. The third is the energy loss $\Delta f^{\mathrm{de}}$ due to the phase separation induced possibly by the spinodal decomposition, strain-controlled boundary conditions, etc. Altogether, the energy difference is

$$\Delta F \leq \gamma_{\mathrm{DI}} S_{\mathrm{DI}} + \frac{1}{2}w^{\mathrm{I}}w^{\mathrm{II}} V[Q(\mathbf{n})+B(\mathbf{n})] - V\Delta f^{\mathrm{de}}, \qquad (7)$$

where $V$ is the volume and $S_{\mathrm{DI}}$ is the area of domain interfaces. The reason for using the less than or equal to symbol $\leq$ instead of the equal symbol $=$ is that the polarization vectors within the two domains would rotate slightly for the incompatible cases to reduce the total energy, which will be discussed in the next section.

From Eq. (7), we may then define a new parameter $\gamma$ as $\gamma = \Delta F / S_{\mathrm{DI}}$, which represents the surface density of the overall energy change caused by the presence of DWs or IBs. For the system in Fig. 1(a), $V = S_{\mathrm{DI}} L / 2$ with $L$ the domain period. Then,

$$\gamma \leq \gamma_{\mathrm{DI}} + \frac{1}{2}\left[\frac{1}{2}w^{\mathrm{I}}w^{\mathrm{II}}[Q(\mathbf{n})+B(\mathbf{n})] - \Delta f^{\mathrm{de}}\right] \cdot L. \qquad (8)$$

Therefore, one can calculate a series of $\gamma$ as a function of the domain period $L$ by numerical approaches such as first-principles calculations, effective Hamiltonian method, phase-field method, etc. The obtained $\gamma$-$L$ curve should be linear. The vertical intercept of this curve represents the short-range interfacial energy $\gamma_{\mathrm{DI}}$, and the double of the slope indicates the long-range contributions from the incompatibilities and phase separations. When the slope of the $\gamma$-$L$ curve is less than zero and the domain period $L$ is larger than a critical value, the coexistence of two domains is energetically more stable compared to the case of mono-domain.

Meanwhile, the critical domain period $L_{\text{crit}}$ can be calculated by solving $\gamma = 0$, which gives

$$L_{\text{crit}} = \frac{2\gamma_{\text{DI}}}{\Delta f^{\text{de}} - \frac{1}{2} w^{\text{I}} w^{\text{II}} \left[ Q(\mathbf{n}) + B(\mathbf{n}) \right]}. \qquad (9)$$

Here $L_{\text{crit}}$ represents the critical length of the domain period when the polydomain structures possess the same energy as the mono-domain ones, which is analogous to the critical radius in the classical nucleation theory [38,39]. It should be noted that $L_{\text{crit}}$ accounts for the gradient energy and represents the balance between the short-range energy terms and the long-range ones. Since the focus of this work is the compatibility of the domain interfaces, the analytical calculations on the interfacial energy $\gamma_{\text{DI}}$ are not included. The critical domain period $L_{\text{crit}}$ can be obtained from the phase-field simulations.

## III. Numerical results from analytical calculations and phase-field simulations

Section II gives a general model that can be used to all FE DWs and IBs. In this section, we apply this model to PbZr$_{0.5}$Ti$_{0.5}$O$_3$ (PZT50) and BiFeO$_3$ (BFO) as two examples. PZT is a perovskite FE material with a remarkable piezoelectric coefficient near the morphotropic phase boundary. The tetragonal (T) and rhombohedral (R) phases coexist in PZT bulks at room temperature, as sketched in Figs. 1(b) and 1(c), respectively.

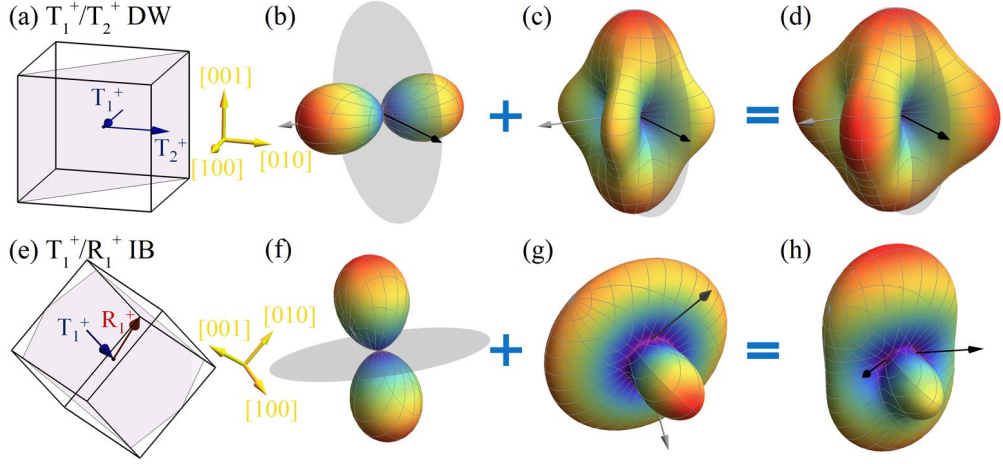

FIG. 2. (a) The sketch of $T_1^+/T_2^+$ domain wall in $PbZr_{0.5}Ti_{0.5}O_3$ (PZT50). (b-d) Polar plots of electrical incompatibility factor $Q(\mathbf{n})$ (b), mechanical incompatibility factor $B(\mathbf{n})$ (c), and the sum of two incompatibility factors $E(\mathbf{n})$ (d) versus the interface orientation $\mathbf{n}$ of $T_1^+/T_2^+$ domain wall. (e) The sketch of $T_1^+/R_1^+$ interphase boundary in PZT50. (f-h) Polar plots of $Q(\mathbf{n})$ (f), $B(\mathbf{n})$ (g), and $E(\mathbf{n})$ (h) versus $\mathbf{n}$ of $T_1^+/R_1^+$ interphase boundary. The shaded planes in (a) and (e) label the interfaces with the lowest $E(\mathbf{n})$. The gray planes in (b) and (f) indicate the possible interface normal with the lowest $Q(\mathbf{n})$. The gray arrows in (c) and (g) indicate the interface normal with the lowest $B(\mathbf{n})$. The black arrows in (d) and (h) indicate the interface normal with the lowest $E(\mathbf{n})$.

According to Eq. (7), we know that only the incompatible energies are related to the orientations of domain interfaces. By minimizing $E(\mathbf{n}) = Q(\mathbf{n}) + B(\mathbf{n})$, one can obtain the interface orientation with the smallest electrostatic and elastic energies. Note that $E(\mathbf{n}) \geq 0$, and the equal sign applies when and only when the interface satisfies both the electrical and mechanical compatibility conditions. Figure 2 shows an example of PZT50 bulks under stress-free boundary conditions at temperature $T = 300K$, in which case the T/T and R/R DWs and the T/R IBs can exist simultaneously (see Section SV of the Supplemental Material [26] for the energy model and corresponding coefficients of PZT50). For the DW between the $T_1^+$ (polarization along [100]) and $T_2^+$ (polarization along [010]) domains [Fig. 2(a)], the interface normal needs to lie within the plane (1-10) to satisfy the electrical compatibility condition, i.e., $\mathbf{n} \parallel$ (1-10), as shown in Fig. 2(b). Meanwhile, by solving the mechanical compatibility condition Eq. (2) or $B(\mathbf{n}) = 0$, $\mathbf{n}$ should be [110] or [1-10], as shown in Fig. 2(c). Altogether, the $T_1^+/T_2^+$ DW normal $\mathbf{n}$ = [110], indicating the typical 90° DWs [Fig. 2(d)]. The R/T IBs, however, are more complicated since the mechanical compatibility condition has no solutions. As illustrated in Fig. 2(g), the boundaries between the $T_1^+$ and $R_1^+$ (polarization along [111]) domains will always generate nonzero mechanical incompatibility factor. By minimizing $E(\mathbf{n})$, two low-energy orientations ($\mathbf{n}$ = [0.70, −0.22, 0.68] and [0.70, 0.68, −0.22]) are obtained, one of which is plotted in Fig. 2(e). Interestingly, these two interfaces are symmetric with respect to the plane spanned by the $T_1^+$ and $R_1^+$ polarization vectors. In general, for any two domains with known polarizations and eigenstrains, the orientations of low-energy domain interfaces can be calculated in a similar manner. In fact, the formation of tetragonal and rhombohedral polydomain structures in FE epitaxial films was studied by Ouyang *et al*. [40] using similar thermodynamic theory [41,42]. The orientation of the elastically best fitting IBs in PZT, PMN-PT, and BFO heterophase polydomain structures could be analytically solved by minimizing the elastic energy. Their result of a {112} type of DWs is close to ours where the interface normal with respect to the minimization of the elastic incompatibility factor $B(\mathbf{n})$ in PZT is

[0.70, 0.50, 0.50] or [−0.68, 0.52, 0.52], as shown in Fig. 2(g).

Another example for the coexistence of different FE phases is BFO. BFO is one of the mostly studied room-temperature multiferroic materials and possesses the R phase below $T_\text{C} \sim$ 1100 K. Moreover, highly strained BFO thin films can form a super-tetragonal ($T^S$) phase with high $c/a$ ratio and large polarization [43,44], and the coexistence of the $T^S$ and distorted rhombohedral ($R^D$) phases was observed in experiments [45]. The polarization vectors of the $T^S$ and $R^D$ phases are illustrated in Figs. 1(d) and 1(e), respectively. To incorporate the contributions from the short-range interfacial energy and long-range interactions, we perform a series of one-dimensional phase-field simulations on the stress-free BFO bulks and highly strained BFO thin films. Following previous works [37,46,47], the total free energy density contains the volume densities of Landau free energy $f_\text{Land}$, gradient energy $f_\text{grad}$, electrostatic energy $f_\text{elec}$, and elastic energy $f_\text{elas}$. There are multiple sets of existing coefficients for BFO [37,48–50], which describe the behaviors of BFO under different conditions. Here we choose the set of coefficients that could describe both the rhombohedral and the super-tetragonal minima (see Section SVI of the Supplemental Material [26] for the energy expressions and coefficients of BFO). It should be noted that the specific values of order parameters may depend on the choice of coefficients, whereas the lowering of the crystal symmetry, which will be shown below, is insensitive to the material coefficients. The mechanical boundary conditions of the stress-free bulks are $\sigma_{ij} = 0$ ($i, j = 1, 2, 3$), and the ones of thin films are assumed to be $\varepsilon_{11} = \varepsilon_{22} = \varepsilon_s$, $\varepsilon_{12} = 0$, and $\sigma_{33} = \sigma_{13} = \sigma_{23} = 0$, where the misfit strain $\varepsilon_s = -4.3\%$ corresponding to the case of BFO films grown on the (001) LaAlO$_3$ substrates.

The system size is chosen to be $N_x\Delta x \times 1\Delta x \times 1\Delta x$ with $\Delta x = 0.02$nm, and $N_x$ is determined by the domain period $L$ via $L = N_x\Delta x$. First, we employ the analytical solutions to obtain the interface orientation **n** with the lowest energy $E(\mathbf{n})$. In the calculation, the polarizations and eigenstrains of the two domains are chosen to be the values of the corresponding mono-domain structures under the same boundary conditions. Then, the coordinate system of the phase-field simulations for the low-energy domain interfaces is rotated such that $\mathbf{x} \parallel \mathbf{n}$ unless mentioned otherwise.

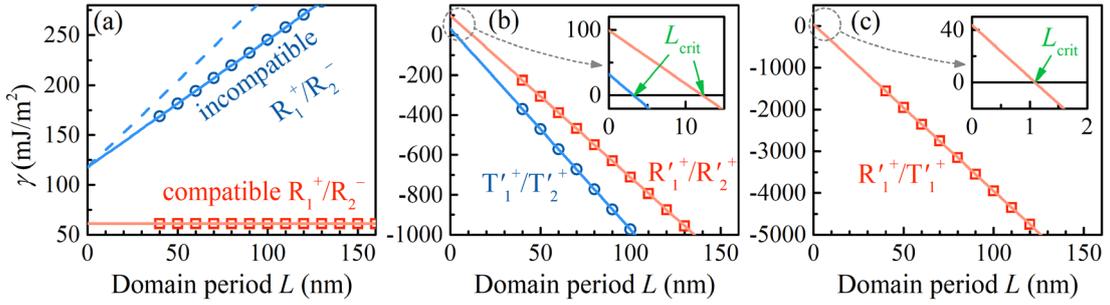

FIG. 3. $\gamma$-$L$ curves of different domain interfaces in stress-free BFO bulks and BFO thin films with misfit strain $\varepsilon_s = -4.3\%$. (a) Compatible and incompatible $R_1^+/R_2^-$ domain walls in stress-free BFO bulks with $\mathbf{n} = [100]$ and $\mathbf{n} = [\sqrt{2}1\text{-}1]$, respectively. (b) $R'^+_1/R'^+_2$ and $T'^+_1/T'^+_2$ domain walls in highly strained BFO thin films. (c) $R'^+_1/T'^+_1$ interphase boundaries in highly strained BFO thin films. Dots represent the results from phase-field simulations, and solid lines are fitted to the dots. The dashed line in (a) is the result from thermodynamic analysis. Insets in (b) and (c) are the zoom-in plots of the grey circular regions showing the critical domain period $L_{\text{crit}}$.

Using phase-field simulations, we calculate $\gamma$ as a function of the domain periods $L$ and plot the $\gamma$-$L$ curves. For the $R_1^+/R_2^-$ (109°) DWs, the value of $\gamma$ is a constant if the interface orientation is chosen to be compatible ($\mathbf{n}$ = [100]), as shown in Fig. 3(a). The zero slope indicates that both the electrical incompatibility factor and the mechanical incompatibility factor are zero. However, if we rotate the interface normal, e.g., from [100] to [$\sqrt{2}$1-1], suggesting that the mechanical compatibility condition is not satisfied, the slope becomes positive, which indicates that the total system energy increases as a function of the domain size. Apparently, such interface orientation is energetically unfavored. In addition, the vertical intercepts of the two lines in Fig. 3(a) are different, meaning that the interfacial energies of the two cases are different.

For BFO thin films, the strain phase separation phenomenon should also be considered. Figure 3(b) shows the results for the $R'^+_1/R'^+_2$ and $T'^+_1/T'^+_2$ DWs, the interface orientations of which both satisfy $E(\mathbf{n}) = 0$. The negative slopes of the two lines result from the strain-induced phase separation process [36,37]. The curve for the $R'^+_1/T'^+_1$ IB is plotted in Fig. 3(c). Since the two compatibility conditions have no solutions for IBs simultaneously, the interface orientation is chosen by minimizing $E(\mathbf{n})$. The slope is also negative, suggesting that the contribution from the phase separation phenomenon dominates over those from the electrostatic and elastic incompatibility energies. Notably, the negative slope of the $\gamma$-$L$ curves, implies the decrease of the system energy caused by the formation of DWs or IBs when the domain period is large enough. In this case, the formation of these interfaces and their symmetric variants is a spontaneous process and thus can be observed in experiments [45].

Moreover, the critical domain period $L_{\text{crit}}$ for a stable domain interface can be determined by solving $\gamma = 0$, as illustrated in the insets of Figs. 3(b) and 3(c).

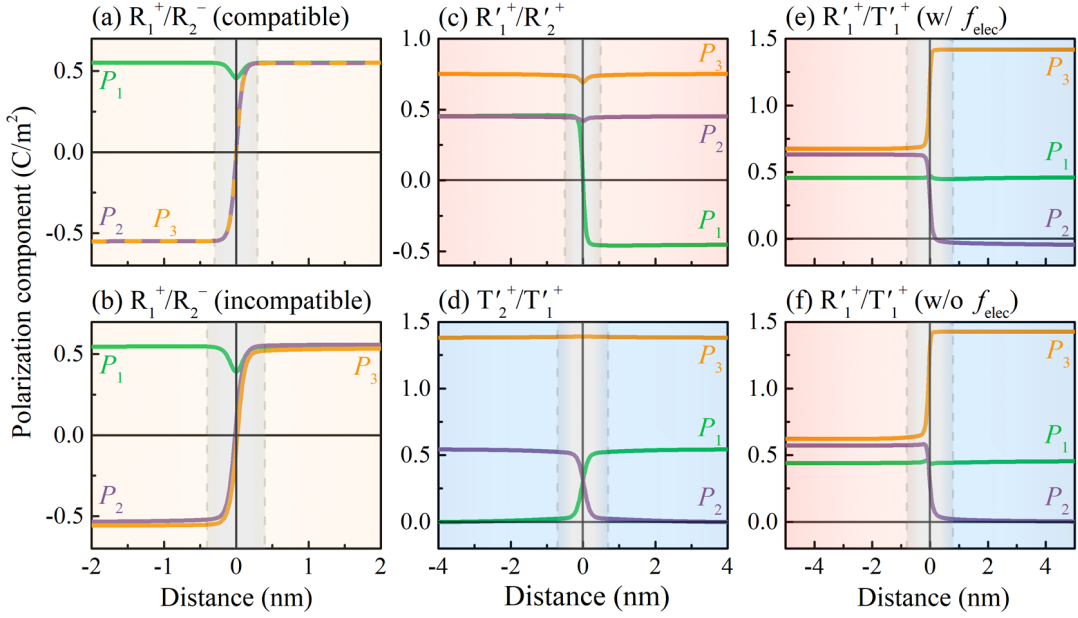

FIG. 4. The polarization profiles across different domain interfaces in stress-free BFO bulks and BFO thin films with misfit strain $\varepsilon_s = -4.3\%$. (a-b) $R_1^+/R_2^-$ domain walls of compatible and incompatible cases with $\mathbf{n} = [100]$ and $\mathbf{n} = [\sqrt{2}1\text{-}1]$, respectively, in stress-free BFO bulks. (c) $R'^+_1/R'^+_2$ domain wall in BFO thin films. (d) $T'^+_1/T'^+_2$ domain wall in BFO thin films. (e-f) $R'^+_1/T'^+_1$ interphase boundaries in BFO thin films including and excluding the electrostatic energy, respectively. The grey shaded regions indicate the interface regions.

From the simulation results, we observe the change of the crystal symmetry within the domains when the interface is an incompatible one. The blue dashed line in Fig. 3(a) is the analytical result whose slope is calculated numerically by the incompatibility factors and is larger than that fitted to the phase-field simulation results [the blue solid line in Fig. 3(a)]. The decrease of the slope of the $\gamma$-$L$ curves could be explained by examining the polarization profiles across the DWs from phase-field simulations, as shown in Figs. 4(a) and 4(b). For the compatible $R_1^+/R_2^-$ DW in Fig. 4(a), the two adjacent domains maintain the same rhombohedral symmetry ($|P_1|=|P_2|=|P_3|$) except for the transitional region in the vicinity of the DW. For the mechanically incompatible case, however, the polarization vectors in both domains rotate slightly away from the <111> directions ($|P_1|\neq|P_2|\neq|P_3|$), as shown in Fig. 4(b). According to Eq. (8), the double of the slope equals to the total excess energy density excluding the local interfacial energy. Therefore, although such polarization rotations increase the Landau free energy density, the elastic energy density can be lowered by reducing the value of the mechanical incompatibility factor. Our calculations shows that the slope of results from phase-field simulations reduces by 26% compared to that calculated from analytical solutions.

Figures 4(c) and 4(d) are the polarization profiles across the $R'^+_1/R'^+_2$ and $T'^+_1/T'^+_2$ DWs in BFO thin films subject to biaxial misfit strain $\varepsilon_s = -4.3\%$, respectively. Similar to the compatible $R_1^+/R_2^-$ DW, the monoclinic symmetry is maintained far away from the DWs. Nevertheless, for the case of the $R'^+_1/T'^+_1$ IB [Fig. 4(e)], the polarization vectors of both the $T^S$ phase and the $R^D$ phase rotate slightly from the directions with monoclinic symmetry, resulting in symmetry breaking (both from monoclinic to triclinic). Meanwhile, the sum of incompatibility factors $E(\mathbf{n})$ reduces by 14% from 3.00 to 2.58 (unit: $10^8$J/m$^3$) caused by the polarization rotation. Note that if the electrostatic energy $f_{\text{elec}}$ is neglected and the interface orientation is chosen by minimizing $B(\mathbf{n})$ in the phase-field simulation [Fig. 4(f)], the polarization rotation angle would become smaller, indicating that the electrical incompatibility is a major factor for the lowering of crystal symmetry in the incompatible polydomain system. Because of the long-range feature of the incompatibility factors, the triclinic phases are also long-range instead of localized near the IBs.

From the above discussion, the crystal symmetry within the domains across compatible DWs ($R^D/R^D$ or $T^S/T^S$) remains unchanged, whereas that across the incompatible IBs ($R^D/T^S$) would be lowered. This finding is supported by earlier experiments where the coexistence of the tilted $R^D$ and tilted $T^S$ phases with triclinic symmetry is observed in BFO thin films grown on the LaAlO$_3$ substrates [45,51–53]. Notably, the origin of the triclinic phases in BFO thin films is still under debate. In previous first-principles calculations on BFO [54], the lowest symmetry of the structures that are stable or metastable under stress-free conditions is the monoclinic symmetry. Our phase-field simulation results demonstrate that those experimentally observed triclinic phases are not intrinsic but instead distorted from monoclinic phases by the long-range electrostatic and elastic energies. In fact, the low-symmetry crystal structures have been found experimentally in other heterophase FE systems such as near the morphotropic phase boundaries in PZT [55,56] and PMN-PT [57] and near the thermotropic phase boundaries in BaTiO$_3$ [58]. In general, the low-energy monoclinic phases may be induced by the substrate constraints and the oxygen octahedral tilts [59] for the single domain case, while the incompatible IBs can also stabilize monoclinic phases in the polydomain systems. The earlier theoretical analysis [60,61] indicates that the crystal lattice of R/T polydomain in FE solid solutions is monoclinic averagely, and the simulations by Ke *et al.* [62,63] suggest that such monoclinic phase near the morphotropic phase boundary is stabilized by the long-range elastic and electrostatic interactions, which is consistent with our simulation results.

## IV. Conclusions

In conclusion, we developed a thermodynamic model to analyze the stability and low-energy orientations of a general domain interface in FE materials, which is applicable to homophase DWs as well as heterophase IBs. We defined the electrical and mechanical incompatibility factors to measure the degrees of electrical and mechanical incompatibilities, respectively. The low-energy orientations of a DW or IB can be predicted by the minimization of the summation of the two factors. Taking PZT50 and BFO as two exemplary systems, we evaluated the low-energy orientations of the DWs and IBs. The competition between the short-range and long-range interactions plays an essential role in the stability of polydomain structures and is demonstrated by phase-field simulations. By comparing the polarization profiles of compatible and incompatible domain interfaces, we found that the electrical and/or mechanical incompatibilities are responsible for the lowering of crystal symmetry in polyphase ferroelectrics. This work provides a general theoretical framework for analyzing the stability and orientations of IBs in FE materials, which can be generalized to other ferroic and multiferroic systems by incorporating other energy terms.

**Acknowledgements**

Y.Z., S.D., and J.M.L. acknowledge the financial support from the National Natural Science Foundation of China (No. 11834002). The phase-field simulations were performed using the commercial software package µ-PRO (http://mupro.co).

**Figure Captions**

FIG. 1. (a) The sketch of the studied system with alternating two domains. (b-c) Polarization directions of the domain variants in lead zirconate titanate (PZT) and bismuth ferrite (BFO) single crystals, including the tetragonal (T) phase and the rhombohedral (R) phase. (d-e) Polarization directions of the domain variants in BFO thin films, including the super-tetragonal ($T^S$) phase and the distorted rhombohedral ($R^D$) phase.

FIG. 2. (a) The sketch of $T_1^+/T_2^+$ domain wall in $PbZr_{0.5}Ti_{0.5}O_3$ (PZT50). (b-d) Polar plots of electrical incompatibility factor $Q(\mathbf{n})$ (b), mechanical incompatibility factor $B(\mathbf{n})$ (c), and the sum of two incompatibility factors $E(\mathbf{n})$ (d) versus the interface orientation $\mathbf{n}$ of $T_1^+/T_2^+$ domain wall. (e) The sketch of $T_1^+/R_1^+$ interphase boundary in PZT50. (f-h) Polar plots of $Q(\mathbf{n})$ (f), $B(\mathbf{n})$ (g), and $E(\mathbf{n})$ (h) versus $\mathbf{n}$ of $T_1^+/R_1^+$ interphase boundary. The shaded planes in (a) and (e) label the interfaces with the lowest $E(\mathbf{n})$. The gray planes in (b) and (f) indicate the possible interface normal with the lowest $Q(\mathbf{n})$. The gray arrows in (c) and (g) indicate the interface normal with the lowest $B(\mathbf{n})$. The black arrows in (d) and (h) indicate the interface normal with the lowest $E(\mathbf{n})$.

FIG. 3. $\gamma$-$L$ curves of different domain interfaces in stress-free BFO bulks and BFO thin films with misfit strain $\varepsilon_s = -4.3\%$. (a) Compatible and incompatible $R_1^+/R_2^-$ domain walls in stress-

free BFO bulks with **n** = [100] and **n** = [$\sqrt{21}$-1], respectively. (b) $R'^+_1/R'^+_2$ and $T'^+_1/T'^+_2$ domain walls in highly strained BFO thin films. (c) $R'^+_1/T'^+_1$ interphase boundaries in highly strained BFO thin films. Dots represent the results from phase-field simulations, and solid lines are fitted to the dots. The dashed line in (a) is the result from thermodynamic analysis. Insets in (b) and (c) are the zoom-in plots of the grey circular regions showing the critical domain period $L_{crit}$.

FIG. 4. The polarization profiles across different domain interfaces in stress-free BFO bulks and BFO thin films with misfit strain $\varepsilon_s$ = –4.3%. (a-b) $R^+_1/R^-_2$ domain walls of compatible and incompatible cases with **n** = [100] and **n** = [$\sqrt{21}$-1], respectively, in stress-free BFO bulks. (c) $R'^+_1/R'^+_2$ domain wall in BFO thin films. (d) $T'^+_1/T'^+_2$ domain wall in BFO thin films. (e-f) $R'^+_1/T'^+_1$ interphase boundaries in BFO thin films including and excluding the electrostatic energy, respectively. The grey shaded regions indicate the interface regions.